\documentclass{article}
\usepackage{spconf}
\usepackage{cite}
\usepackage[T1]{fontenc}
\usepackage{graphicx}
\usepackage{amssymb}
\usepackage{bbm}
\usepackage{amsmath}
\usepackage{amsthm}
\usepackage{subcaption}
\usepackage{microtype}
\usepackage{balance}
\usepackage{xcolor}
\usepackage{algorithm}
\usepackage{algorithmicx}
\usepackage{algpseudocode}
\usepackage[acronym]{glossaries}
\usepackage{url}
\usepackage{float}
\usepackage{booktabs}
\usepackage{tabularx}
\usepackage{hyperref}

\algrenewcommand\algorithmicindent{0.7em}%

\newacronym{NPRACH}{NPRACH}{narrowband physical random-access channel}
\newacronym{ToA}{ToA}{time of arrival}
\newacronym{CFO}{CFO}{carrier frequency offset}
\newacronym{5GNR}{5G NR}{5G New Radio}
\newacronym{3GPP}{3GPP}{3rd Generation Partnership Project}
\newacronym{UMi}{UMi}{urban microcell}
\newacronym{RMSE}{RMSE}{root-mean-square error}
\newacronym{NN}{NN}{neural network}
\newacronym{BS}{BS}{base station}
\newacronym{UE}{UE}{user equipment}
\newacronym{CP}{CP}{cyclic prefix}
\newacronym{OFDM}{OFDM}{orthogonal frequency division multiplexing}
\newacronym{FFT}{FFT}{fast Fourier transform}
\newacronym{AWGN}{AWGN}{additive white Gaussian noise}
\newacronym{DFT}{DFT}{discrete Fourier transform}
\newacronym{RG}{RG}{resource grid}
\newacronym{RE}{RE}{resource element}
\newacronym{SNR}{SNR}{signal-to-noise ratio}
\newacronym{SINR}{SINR}{signal-to-interference-plus-noise ratio}
\newacronym{MLP}{MLP}{multilayer perceptron}
\newacronym{BCE}{BCE}{binary cross-entropy}
\newacronym{CCE}{CCE}{categorical cross-entropy}
\newacronym{ERM}{ERM}{empirical risk minimization}
\newacronym{MC}{MC}{Monte Carlo}
\newacronym{KL}{KL}{Kullback–Leibler}
\newacronym{SGD}{SGD}{stochastic gradient descent}
\newacronym{ICI}{ICI}{inter-carrier interference}
\newacronym{GNN}{GNN}{graph neural network}
\newacronym{BP}{BP}{belief propagation}
\newacronym{MP}{MP}{message passing}
\newacronym{FEC}{FEC}{forward error correction}
\newacronym{LDPC}{LDPC}{low-density parity-check}
\newacronym{IQ}{I/Q}{in-phase/quadrature}
\newacronym{HDPC}{HDPC}{high-density parity-check}
\newacronym{SCL}{SCL}{successive cancellation list}
\newacronym{SC}{SC}{successive cancellation}
\newacronym{URLLC}{URLLC}{ultra-reliable low-latency communications}
\newacronym{APP}{APP}{a posterior probability}
\newacronym{MIMO}{MIMO}{multiple-input multiple-output}
\newacronym{CNN}{CNN}{convolutional neural network}
\newacronym{ResNet}{ResNet}{residual network}
\newacronym{BER}{BER}{bit error rate}
\newacronym{BPSK}{BPSK}{binary phase shift keying}
\newacronym{LLR}{LLR}{log-likelihood ratio}
\newacronym{VN}{VN}{variable node}
\newacronym{CN}{CN}{check node}
\newacronym{MPNN}{MPNN}{message passing neural network}

\newacronym{AI}{AI}{artificial intelligence}
\newacronym{ML}{ML}{machine learning}
\newacronym{LLM}{LLM}{large language model}
\newacronym{SISO}{SISO}{single-input single-output; soft-input soft-output}
\newacronym{MISO}{MISO}{multiple-input single-output}
\newacronym{PRB}{PRB}{physical resource block}
\newacronym{PUSCH}{PUSCH}{physical uplink shared channel}
\newacronym{PDSCH}{PDSCH}{physical downlink shared channel}
\newacronym{PUCCH}{PUCCH}{physical uplink control channel}
\newacronym{PDCCH}{PDCCH}{physical downlink control channel}
\newacronym{PRACH}{PRACH}{physical random-access channel}
\newacronym{SRS}{SRS}{sounding reference signal}
\newacronym{PBCH}{PBCH}{physical broadcast channel}
\newacronym{SSB}{SSB}{synchronization signal block}
\newacronym{AMF}{AMF}{access and mobility management function}
\newacronym{SMF}{SMF}{session management function}
\newacronym{PDU}{PDU}{protocol data unit}
\newacronym{NGAP}{NGAP}{next generation application protocol}
\newacronym{IMS}{IMS}{IP multimedia subsystem}
\newacronym{IP}{IP}{Internet Protocol}
\newacronym{SIP}{SIP}{session initiation protocol}
\newacronym{DN}{DN}{data network}
\newacronym{UPF}{UPF}{user plane function}
\newacronym{AUSF}{AUSF}{authentication server function}
\newacronym{UDM}{UDM}{unified data management}
\newacronym{UDR}{UDR}{unified data repository}
\newacronym{NRF}{NRF}{network repository function}
\newacronym{HARQ}{HARQ}{hybrid automatic repeat request}
\newacronym{LA}{LA}{link adaptation}
\newacronym{SE}{SE}{spectral efficiency}
\newacronym{TCP}{TCP}{transmission control protocol}
\newacronym{API}{API}{application programming interface}
\newacronym{SQL}{SQL}{structured query language}
\newacronym{ARF}{ARF}{auto rate fallback}
\newacronym{ACK}{ACK}{acknowledgment}
\newacronym{NACK}{NACK}{negative acknowledgment}
\newacronym{OLLA}{OLLA}{outer-loop link adaptation}
\newacronym{CQI}{CQI}{channel quality indicator}
\newacronym{RL}{RL}{reinforcement learning}
\newacronym{MAB}{MAB}{multi-armed bandit}
\newacronym{SALAD}{SALAD}{self-adaptive link adaptation}
\newacronym{ILLA}{ILLA}{inner-loop link adaptation}
\newacronym{OGD}{OGD}{online gradient descent}

\newacronym{MUMIMO}{MU-MIMO}{multi-user multiple-input multiple-output}
\newacronym{ULL}{ULL}{uplink layer}
\newacronym{BICM}{BICM}{bit-interleaved coded modulation}
\newacronym{QAM}{QAM}{quadrature amplitude modulation}
\newacronym{LMMSE}{LMMSE}{linear minimum mean square error}
\newacronym{MMSE}{MMSE}{minimum mean square error}
\newacronym{CSI}{CSI}{channel-state information}
\newacronym{SIMO}{SIMO}{single-input multiple-output}
\newacronym{SLAM}{SLAM}{simultaneous localization and mapping}
\newacronym{LIDAR}{LIDAR}{light detection and ranging}
\newacronym{CGNN}{CGNN}{convolutional and graph neural network}
\newacronym{BLER}{BLER}{block error rate}
\newacronym{LS}{LS}{least squares}
\newacronym{PE}{PE}{positional encoding}
\newacronym{relu}{ReLU}{rectified linear unit}
\newacronym{RB}{RB}{resource block}
\newacronym{DMRS}{DMRS}{demodulation reference signal}
\newacronym{CSI-RS}{CSI-RS}{channel state information reference signal}
\newacronym{IoT}{IoT}{internet of things}
\newacronym{ADAM}{ADAM}{adaptive momentum}
\newacronym{TBLER}{TBLER}{transport block error rate}
\newacronym{TBS}{TBS}{transport block size}
\newacronym{MCS}{MCS}{modulation and coding scheme}
\newacronym{TDL}{TDL}{tapped delay line}
\newacronym{CDL}{CDL}{clustered delay line}
\newacronym{GSCM}{GSCM}{geometry-based stochastic channel model}
\newacronym{CDM}{CDM}{code division multiplexing}
\newacronym{FLOP}{FLOP}{floating point operation}
\newacronym{PHY}{PHY}{physical layer}
\newacronym{MAC}{MAC}{media access control}
\newacronym{gNB}{gNB}{next-generation Node B}
\newacronym{RLC}{RLC}{radio link control}
\newacronym{PDCP}{PDCP}{packet data convergence protocol}
\newacronym{SDAP}{SDAP}{service data adaptation protocol}
\newacronym{RRC}{RRC}{radio resource control}

\newacronym{ULA}{ULA}{uniform linear array}
\newacronym{NRX}{NRX}{neural receiver}
\newacronym{MDX}{MDX}{model-driven neural receiver}
\newacronym{GPU}{GPU}{graphics processing unit}
\newacronym{CPU}{CPU}{central processing unit}
\newacronym{NIC}{NIC}{network interface card}
\newacronym{Var-MCS-NRX}{Var-MCS NRX}{variable-MCS NRX}
\newacronym{UL}{UL}{uplink}
\newacronym{DL}{DL}{downlink}
\newacronym{MSE}{MSE}{mean squared error}
\newacronym{CIR}{CIR}{channel impulse response}
\newacronym{iid}{iid}{independent and identically distributed}

\newacronym{RAN}{RAN}{radio access network}
\newacronym{ORU}{O-RU}{open RAN radio unit}
\newacronym{ORAN}{O-RAN}{open radio access network}
\newacronym{COTS}{COTS}{commercial-off-the-shelf}
\newacronym{RF}{RF}{radio frequency}
\newacronym{LOS}{LOS}{line-of-sight}
\newacronym{NLOS}{NLOS}{non-line-of-sight}
\newacronym{OTA}{OTA}{over-the-air}
\newacronym{TDD}{TDD}{time-division duplexing}
\newacronym{CAEZ}{CAEZ}{CSI acquisition at ETH Zurich}
\newacronym{ATB}{ATB}{NVIDIA Aerial Testbed}
\newacronym{ARK}{ARK}{NVIDIA Aerial Research Kit}
\newacronym{OAI}{OAI}{OpenAirInterface}
\newacronym{RFFI}{RFFI}{radio frequency fingerprint identification}
\newacronym{SVD}{SVD}{singular value decomposition}
\newacronym{NLER}{NLER}{neighbor label error rate}
\newacronym{tSNE}{t-SNE}{$t$-distributed stochastic neighbor embedding}
\newacronym{FH}{FH}{front haul}
\newacronym{GNSS}{GNSS}{global navigation satellite system}
\newacronym{RTK}{RTK}{real-time kinematic}
\newacronym{PTP}{PTP}{precision time protocol}
\newacronym{RNTI}{RNTI}{radio network temporary identifier}
\newacronym{ISM}{ISM}{Industrial, Scientific, and Medical}
\newacronym{UAV}{UAV}{unmanned aerial vehicle}
\newacronym{CDF}{CDF}{cumulative distribution function}
\newacronym{CRC}{CRC}{cyclic redundancy check}
\newacronym{TP}{TP}{throughput}

\newacronym{IDD}{IDD}{iterative detection and decoding}
\newacronym{DUIDD}{DUIDD}{deep-unfolded interleaved detection and decoding}
\newacronym{SPAT}{SPAT}{swapping of punctured and transmitted blocks}

\usepackage{amssymb}
\usepackage{amsfonts}
\usepackage{mathrsfs}
\usepackage{xspace}
\usepackage{bm}
\usepackage{upgreek}

\newcommand{\safemath}[2]{\newcommand{#1}{\ensuremath{#2}\xspace}}

\safemath{\bma}{\mathbf{a}}
\safemath{\bmb}{\mathbf{b}}
\safemath{\bmc}{\mathbf{c}}
\safemath{\bmd}{\mathbf{d}}
\safemath{\bme}{\mathbf{e}}
\safemath{\bmf}{\mathbf{f}}
\safemath{\bmg}{\mathbf{g}}
\safemath{\bmh}{\mathbf{h}}
\safemath{\bmi}{\mathbf{i}}
\safemath{\bmj}{\mathbf{j}}
\safemath{\bmk}{\mathbf{k}}
\safemath{\bml}{\mathbf{l}}
\safemath{\bmm}{\mathbf{m}}
\safemath{\bmn}{\mathbf{n}}
\safemath{\bmo}{\mathbf{o}}
\safemath{\bmp}{\mathbf{p}}
\safemath{\bmq}{\mathbf{q}}
\safemath{\bmr}{\mathbf{r}}
\safemath{\bms}{\mathbf{s}}
\safemath{\bmt}{\mathbf{t}}
\safemath{\bmu}{\mathbf{u}}
\safemath{\bmv}{\mathbf{v}}
\safemath{\bmw}{\mathbf{w}}
\safemath{\bmx}{\mathbf{x}}
\safemath{\bmy}{\mathbf{y}}
\safemath{\bmz}{\mathbf{z}}
\safemath{\bmzero}{\mathbf{0}}
\safemath{\bmone}{\mathbf{1}}
\safemath{\Bell}{\ensuremath{\boldsymbol\ell}}

\bmdefine{\biad}{a}
\bmdefine{\bibd}{b}
\bmdefine{\bicd}{c}
\bmdefine{\bidd}{d}
\bmdefine{\bied}{e}
\bmdefine{\bifd}{f}
\bmdefine{\bigd}{g}
\bmdefine{\bihd}{h}
\bmdefine{\biid}{i}
\bmdefine{\bijd}{j}
\bmdefine{\bikd}{k}
\bmdefine{\bild}{l}
\bmdefine{\bimd}{m}
\bmdefine{\bind}{n}
\bmdefine{\biod}{o}
\bmdefine{\bipd}{p}
\bmdefine{\biqd}{q}
\bmdefine{\bird}{r}
\bmdefine{\bisd}{s}
\bmdefine{\bitd}{t}
\bmdefine{\biud}{u}
\bmdefine{\bivd}{v}
\bmdefine{\biwd}{w}
\bmdefine{\bixd}{x}
\bmdefine{\biyd}{y}
\bmdefine{\bizd}{z}

\bmdefine{\bixid}{\xi}
\bmdefine{\bilambdad}{\lambda}
\bmdefine{\bimud}{\mu}
\bmdefine{\bithetad}{\theta}
\bmdefine{\biphid}{\phi}
\bmdefine{\bideltad}{\delta}

\safemath{\bmia}{\biad}
\safemath{\bmib}{\bibd}
\safemath{\bmic}{\bicd}
\safemath{\bmid}{\bidd}
\safemath{\bmie}{\bied}
\safemath{\bmif}{\bifd}
\safemath{\bmig}{\bigd}
\safemath{\bmih}{\bihd}
\safemath{\bmii}{\biid}
\safemath{\bmij}{\bijd}
\safemath{\bmik}{\bikd}
\safemath{\bmil}{\bild}
\safemath{\bmim}{\bimd}
\safemath{\bmin}{\bind}
\safemath{\bmio}{\biod}
\safemath{\bmip}{\bipd}
\safemath{\bmiq}{\biqd}
\safemath{\bmir}{\bird}
\safemath{\bmis}{\bisd}
\safemath{\bmit}{\bitd}
\safemath{\bmiu}{\biud}
\safemath{\bmiv}{\bivd}
\safemath{\bmiw}{\biwd}
\safemath{\bmix}{\bixd}
\safemath{\bmiy}{\biyd}
\safemath{\bmiz}{\bizd}

\safemath{\bmxi}{\bixid}
\safemath{\bmlambda}{\bilambdad}
\safemath{\bmmu}{\bimud}
\safemath{\bmtheta}{\bithetad}
\safemath{\bmphi}{\biphid}
\safemath{\bmdelta}{\bideltad}

\safemath{\bA}{\mathbf{A}}
\safemath{\bB}{\mathbf{B}}
\safemath{\bC}{\mathbf{C}}
\safemath{\bD}{\mathbf{D}}
\safemath{\bE}{\mathbf{E}}
\safemath{\bF}{\mathbf{F}}
\safemath{\bG}{\mathbf{G}}
\safemath{\bH}{\mathbf{H}}
\safemath{\bI}{\mathbf{I}}
\safemath{\bJ}{\mathbf{J}}
\safemath{\bK}{\mathbf{K}}
\safemath{\bL}{\mathbf{L}}
\safemath{\bM}{\mathbf{M}}
\safemath{\bN}{\mathbf{N}}
\safemath{\bO}{\mathbf{O}}
\safemath{\bP}{\mathbf{P}}
\safemath{\bQ}{\mathbf{Q}}
\safemath{\bR}{\mathbf{R}}
\safemath{\bS}{\mathbf{S}}
\safemath{\bT}{\mathbf{T}}
\safemath{\bU}{\mathbf{U}}
\safemath{\bV}{\mathbf{V}}
\safemath{\bW}{\mathbf{W}}
\safemath{\bX}{\mathbf{X}}
\safemath{\bY}{\mathbf{Y}}
\safemath{\bZ}{\mathbf{Z}}

\safemath{\bZero}{\mathbf{0}}
\safemath{\bOne}{\mathbf{1}}
\safemath{\bDelta}{\mathbf{\Delta}}
\safemath{\bLambda}{\mathbf{\UpLambda}}
\safemath{\bPhi}{\mathbf{\Upphi}}
\safemath{\bSigma}{\mathbf{\Upsigma}}
\safemath{\bOmega}{\mathbf{\Upomega}}
\safemath{\bTheta}{\mathbf{\Uptheta}}

\bmdefine{\biAd}{A}
\bmdefine{\biBd}{B}
\bmdefine{\biCd}{C}
\bmdefine{\biDd}{D}
\bmdefine{\biEd}{E}
\bmdefine{\biFd}{F}
\bmdefine{\biGd}{G}
\bmdefine{\biHd}{H}
\bmdefine{\biId}{I}
\bmdefine{\biJd}{J}
\bmdefine{\biKd}{K}
\bmdefine{\biLd}{L}
\bmdefine{\biMd}{M}
\bmdefine{\biOd}{N}
\bmdefine{\biPd}{O}
\bmdefine{\biQd}{P}
\bmdefine{\biRd}{R}
\bmdefine{\biSd}{S}
\bmdefine{\biTd}{T}
\bmdefine{\biUd}{U}
\bmdefine{\biVd}{V}
\bmdefine{\biWd}{W}
\bmdefine{\biXd}{X}
\bmdefine{\biYd}{Y}
\bmdefine{\biZd}{Z}

\bmdefine{\biDelta}{\Delta}
\bmdefine{\biLambda}{\Lambda}
\bmdefine{\biPhi}{\Phi}
\bmdefine{\biSigma}{\Sigma}
\bmdefine{\biOmega}{\Omega}
\bmdefine{\biTheta}{\Theta}

\safemath{\bimA}{\biAd}
\safemath{\bimB}{\biBd}
\safemath{\bimC}{\biCd}
\safemath{\bimD}{\biDd}
\safemath{\bimE}{\biEd}
\safemath{\bimF}{\biFd}
\safemath{\bimG}{\biGd}
\safemath{\bimH}{\biHd}
\safemath{\bimI}{\biId}
\safemath{\bimJ}{\biJd}
\safemath{\bimK}{\biKd}
\safemath{\bimL}{\biLd}
\safemath{\bimM}{\biMd}
\safemath{\bimN}{\biNd}
\safemath{\bimO}{\biOd}
\safemath{\bimP}{\biPd}
\safemath{\bimQ}{\biQd}
\safemath{\bimR}{\biRd}
\safemath{\bimS}{\biSd}
\safemath{\bimT}{\biTd}
\safemath{\bimU}{\biUd}
\safemath{\bimV}{\biVd}
\safemath{\bimW}{\biWd}
\safemath{\bimX}{\biXd}
\safemath{\bimY}{\biYd}
\safemath{\bimZ}{\biZd}

\safemath{\bimDelta}{\biDelta}
\safemath{\bimLambda}{\biLambda}
\safemath{\bimPhi}{\biPhi}
\safemath{\bimSigma}{\biSigma}
\safemath{\bimOmega}{\biOmega}
\safemath{\bimTheta}{\biTheta}

\safemath{\setA}{\mathcal{A}}
\safemath{\setB}{\mathcal{B}}
\safemath{\setC}{\mathcal{C}}
\safemath{\setD}{\mathcal{D}}
\safemath{\setE}{\mathcal{E}}
\safemath{\setF}{\mathcal{F}}
\safemath{\setG}{\mathcal{G}}
\safemath{\setH}{\mathcal{H}}
\safemath{\setI}{\mathcal{I}}
\safemath{\setJ}{\mathcal{J}}
\safemath{\setK}{\mathcal{K}}
\safemath{\setL}{\mathcal{L}}
\safemath{\setM}{\mathcal{M}}
\safemath{\setN}{\mathcal{N}}
\safemath{\setO}{\mathcal{O}}
\safemath{\setP}{\mathcal{P}}
\safemath{\setQ}{\mathcal{Q}}
\safemath{\setR}{\mathcal{R}}
\safemath{\setS}{\mathcal{S}}
\safemath{\setT}{\mathcal{T}}
\safemath{\setU}{\mathcal{U}}
\safemath{\setV}{\mathcal{V}}
\safemath{\setW}{\mathcal{W}}
\safemath{\setX}{\mathcal{X}}
\safemath{\setY}{\mathcal{Y}}
\safemath{\setZ}{\mathcal{Z}}
\safemath{\emptySet}{\varnothing}

\safemath{\colA}{\mathscr{A}}
\safemath{\colB}{\mathscr{B}}
\safemath{\colC}{\mathscr{C}}
\safemath{\colD}{\mathscr{D}}
\safemath{\colE}{\mathscr{E}}
\safemath{\colF}{\mathscr{F}}
\safemath{\colG}{\mathscr{G}}
\safemath{\colH}{\mathscr{H}}
\safemath{\colI}{\mathscr{I}}
\safemath{\colJ}{\mathscr{J}}
\safemath{\colK}{\mathscr{K}}
\safemath{\colL}{\mathscr{L}}
\safemath{\colM}{\mathscr{M}}
\safemath{\colN}{\mathscr{N}}
\safemath{\colO}{\mathscr{O}}
\safemath{\colP}{\mathscr{P}}
\safemath{\colQ}{\mathscr{Q}}
\safemath{\colR}{\mathscr{R}}
\safemath{\colS}{\mathscr{S}}
\safemath{\colT}{\mathscr{T}}
\safemath{\colU}{\mathscr{U}}
\safemath{\colV}{\mathscr{V}}
\safemath{\colW}{\mathscr{W}}
\safemath{\colX}{\mathscr{X}}
\safemath{\colY}{\mathscr{Y}}
\safemath{\colZ}{\mathscr{Z}}

\safemath{\opA}{\mathbb{A}}
\safemath{\opB}{\mathbb{B}}
\safemath{\opC}{\mathbb{C}}
\safemath{\opD}{\mathbb{D}}
\safemath{\opE}{\mathbb{E}}
\safemath{\opF}{\mathbb{F}}
\safemath{\opG}{\mathbb{G}}
\safemath{\opH}{\mathbb{H}}
\safemath{\opI}{\mathbb{I}}
\safemath{\opJ}{\mathbb{J}}
\safemath{\opK}{\mathbb{K}}
\safemath{\opL}{\mathbb{L}}
\safemath{\opM}{\mathbb{M}}
\safemath{\opN}{\mathbb{N}}
\safemath{\opO}{\mathbb{O}}
\safemath{\opP}{\mathbb{P}}
\safemath{\opQ}{\mathbb{Q}}
\safemath{\opR}{\mathbb{R}}
\safemath{\opS}{\mathbb{S}}
\safemath{\opT}{\mathbb{T}}
\safemath{\opU}{\mathbb{U}}
\safemath{\opV}{\mathbb{V}}
\safemath{\opW}{\mathbb{W}}
\safemath{\opX}{\mathbb{X}}
\safemath{\opY}{\mathbb{Y}}
\safemath{\opZ}{\mathbb{Z}}
\safemath{\opZero}{\mathbb{O}}
\safemath{\identityop}{\opI}

\safemath{\veca}{\bma}
\safemath{\vecb}{\bmb}
\safemath{\vecc}{\bmc}
\safemath{\vecd}{\bmd}
\safemath{\vece}{\bme}
\safemath{\vecf}{\bmf}
\safemath{\vecg}{\bmg}
\safemath{\vech}{\bmh}
\safemath{\veci}{\bmi}
\safemath{\vecj}{\bmj}
\safemath{\veck}{\bmk}
\safemath{\vecl}{\bml}
\safemath{\vecm}{\bmm}
\safemath{\vecn}{\bmn}
\safemath{\veco}{\bmo}
\safemath{\vecp}{\bmp}
\safemath{\vecq}{\bmq}
\safemath{\vecr}{\bmr}
\safemath{\vecs}{\bms}
\safemath{\vect}{\bmt}
\safemath{\vecu}{\bmu}
\safemath{\vecv}{\bmv}
\safemath{\vecw}{\bmw}
\safemath{\vecx}{\bmx}
\safemath{\vecy}{\bmy}
\safemath{\vecz}{\bmz}

\safemath{\veczero}{\bmzero}
\safemath{\vecone}{\bmone}
\safemath{\vecxi}{\bmxi}
\safemath{\veclambda}{\bmlambda}
\safemath{\vecmu}{\bmmu}
\safemath{\vectheta}{\bmtheta}
\safemath{\vecphi}{\bmphi}
\safemath{\vecdelta}{\bmdelta}

\safemath{\matA}{\bA}
\safemath{\matB}{\bB}
\safemath{\matC}{\bC}
\safemath{\matD}{\bD}
\safemath{\matE}{\bE}
\safemath{\matF}{\bF}
\safemath{\matG}{\bG}
\safemath{\matH}{\bH}
\safemath{\matI}{\bI}
\safemath{\matJ}{\bJ}
\safemath{\matK}{\bK}
\safemath{\matL}{\bL}
\safemath{\matM}{\bM}
\safemath{\matN}{\bN}
\safemath{\matO}{\bO}
\safemath{\matP}{\bP}
\safemath{\matQ}{\bQ}
\safemath{\matR}{\bR}
\safemath{\matS}{\bS}
\safemath{\matT}{\bT}
\safemath{\matU}{\bU}
\safemath{\matV}{\bV}
\safemath{\matW}{\bW}
\safemath{\matX}{\bX}
\safemath{\matY}{\bY}
\safemath{\matZ}{\bZ}
\safemath{\matzero}{\bmzero}

\safemath{\matDelta}{\bDelta}
\safemath{\matLambda}{\bLambda}
\safemath{\matPhi}{\bPhi}
\safemath{\matSigma}{\bSigma}
\safemath{\matOmega}{\bOmega}
\safemath{\matTheta}{\bTheta}

\safemath{\matidentity}{\matI}
\safemath{\matone}{\matO}

\safemath{\rnda}{A}
\safemath{\rndb}{B}
\safemath{\rndc}{C}
\safemath{\rndd}{D}
\safemath{\rnde}{E}
\safemath{\rndf}{F}
\safemath{\rndg}{G}
\safemath{\rndh}{H}
\safemath{\rndi}{I}
\safemath{\rndj}{J}
\safemath{\rndk}{K}
\safemath{\rndl}{L}
\safemath{\rndm}{M}
\safemath{\rndn}{N}
\safemath{\rndo}{O}
\safemath{\rndp}{P}
\safemath{\rndq}{Q}
\safemath{\rndr}{R}
\safemath{\rnds}{S}
\safemath{\rndt}{T}
\safemath{\rndu}{U}
\safemath{\rndv}{V}
\safemath{\rndw}{W}
\safemath{\rndx}{X}
\safemath{\rndy}{Y}
\safemath{\rndz}{Z}

\safemath{\rveca}{\bimA}
\safemath{\rvecb}{\bimB}
\safemath{\rvecc}{\bimC}
\safemath{\rvecd}{\bimD}
\safemath{\rvece}{\bimE}
\safemath{\rvecf}{\bimF}
\safemath{\rvecg}{\bimG}
\safemath{\rvech}{\bimH}
\safemath{\rveci}{\bimI}
\safemath{\rvecj}{\bimJ}
\safemath{\rveck}{\bimK}
\safemath{\rvecl}{\bimL}
\safemath{\rvecm}{\bimM}
\safemath{\rvecn}{\bimN}
\safemath{\rveco}{\bomO}
\safemath{\rvecp}{\bimP}
\safemath{\rvecq}{\bimQ}
\safemath{\rvecr}{\bimR}
\safemath{\rvecs}{\bimS}
\safemath{\rvect}{\bimT}
\safemath{\rvecu}{\bimU}
\safemath{\rvecv}{\bimV}
\safemath{\rvecw}{\bimW}
\safemath{\rvecx}{\bimX}
\safemath{\rvecy}{\bimY}
\safemath{\rvecz}{\bimZ}

\safemath{\rvecxi}{\bmxi}
\safemath{\rveclambda}{\bmlambda}
\safemath{\rvecmu}{\bmmu}
\safemath{\rvectheta}{\bmtheta}
\safemath{\rvecphi}{\bmphi}

\safemath{\rmatA}{\bimA}
\safemath{\rmatB}{\bimB}
\safemath{\rmatC}{\bimC}
\safemath{\rmatD}{\bimD}
\safemath{\rmatE}{\bimE}
\safemath{\rmatF}{\bimF}
\safemath{\rmatG}{\bimG}
\safemath{\rmatH}{\bimH}
\safemath{\rmatI}{\bimI}
\safemath{\rmatJ}{\bimJ}
\safemath{\rmatK}{\bimK}
\safemath{\rmatL}{\bimL}
\safemath{\rmatM}{\bimM}
\safemath{\rmatN}{\bimN}
\safemath{\rmatO}{\bimO}
\safemath{\rmatP}{\bimP}
\safemath{\rmatQ}{\bimQ}
\safemath{\rmatR}{\bimR}
\safemath{\rmatS}{\bimS}
\safemath{\rmatT}{\bimT}
\safemath{\rmatU}{\bimU}
\safemath{\rmatV}{\bimV}
\safemath{\rmatW}{\bimW}
\safemath{\rmatX}{\bimX}
\safemath{\rmatY}{\bimY}
\safemath{\rmatZ}{\bimZ}

\safemath{\rmatDelta}{\bimDelta}
\safemath{\rmatLambda}{\bimLambda}
\safemath{\rmatPhi}{\bimPhi}
\safemath{\rmatSigma}{\bimSigma}
\safemath{\rmatOmega}{\bimOmega}
\safemath{\rmatTheta}{\bimTheta}

\usepackage{amssymb}
\usepackage{amsfonts}
\usepackage{mathrsfs}
\usepackage{xspace}
\usepackage{bm}
\usepackage{fancyref}
\usepackage{textcomp}

\usepackage{multirow}
\usepackage{stmaryrd}

\newenvironment{textbmatrix}{	\setlength{\arraycolsep}{2.5pt}%
								\left[\begin{matrix}}{\end{matrix}\right]%
								\raisebox{0.08ex}{\vphantom{M}}}

\def\be{\begin{equation}}
\def\ee{\end{equation}}
\def\een{\nonumber \end{equation}}
\def\mat{\begin{bmatrix}}
\def\emat{\end{bmatrix}}
\def\btm{\begin{textbmatrix}}
\def\etm{\end{textbmatrix}}

\def\ba#1\ea{\begin{align}#1\end{align}}
\def\bas#1\eas{\begin{align*}#1\end{align*}}
\def\bs#1\es{\begin{split}#1\end{split}}
\def\bg#1\eg{\begin{gather}#1\end{gather}}
\def\bml#1\eml{\begin{multline}#1\end{multline}}
\def\bi#1\ei{\begin{itemize}#1\end{itemize}}

\safemath{\dirac}{\delta}					% Dirac delta
\safemath{\krond}{\dirac}					% Kronecker delta
\safemath{\upto}{\uparrow}
\safemath{\downto}{\downarrow}
\safemath{\iu}{j}							% imaginary unit
\safemath{\ev}{\lambda}						% eigenvalue
\safemath{\hilseqspace}{l^{2}}				% Hilbert sequence space
\newcommand{\banachfunspace}[1]{\setL^{#1}}	% Banach function space
\safemath{\hilfunspace}{\banachfunspace{2}}	% Hilbert function space
\safemath{\SNR}{\textit{SNR}} 				% signal to noise ratio
\safemath{\PAR}{\textit{PAR}} 				% signal to noise ratio
\safemath{\No}{N_0}							% noise spectral density
\safemath{\Es}{E_s}							% energy per symbol
\safemath{\Eb}{E_b}							% energy per bit
\safemath{\EbNo}{\frac{\Eb}{\No}}
\safemath{\EsNo}{\frac{\Es}{\No}}

\DeclareMathOperator{\CHop}{\ensuremath{\opH}} % channel operator
\safemath{\tvir}{\rndh_{\CHop}}				% time-varying impulse response
\safemath{\tvtf}{\rndl_{\CHop}}				% 	-''- transfer function
\safemath{\spf}{\rnds_{\CHop}}				% spreading function
\safemath{\bff}{H_{\CHop}}					% bi-freuqency function

\safemath{\ircf}{r_{h}}						% impulse response correlation fn.
\safemath{\tftvcf}{r_{s}}					% scattering function
\safemath{\tfcf}{r_{l}}						% time-frequency correlation fn.
\safemath{\bfcf}{r_{H}}						% bi-frequency correlation fn.

\safemath{\tcorr}{c_h}						% time-correlation function
\safemath{\scf}{c_{s}}						% spreading function
\safemath{\tfcorr}{c_{l}}					% transfer-function correlation
\safemath{\fcorr}{c_{H}}						% frequency-correlation function

\safemath{\mi}{I}							% mutual information
\safemath{\capacity}{C}						% capacity

\safemath{\normal}{\mathcal{N}}			% normal distribution
\safemath{\jpg}{\mathcal{CN}}			% jointly proper Gaussian
\safemath{\mchain}{\leftrightarrow}		% Markov chain
\safemath{\dB}{\,\mathrm{dB}}
\safemath{\dBm}{\,\mathrm{dBm}}
\safemath{\Hz}{\,\mathrm{Hz}}
\safemath{\kHz}{\,\mathrm{kHz}}
\safemath{\MHz}{\,\mathrm{MHz}}
\safemath{\GHz}{\,\mathrm{GHz}}
\safemath{\s}{\,\mathrm{s}}
\safemath{\ms}{\,\mathrm{ms}}
\safemath{\mus}{\,\mathrm{\text{\textmu}s}}
\safemath{\ns}{\,\mathrm{ns}}
\safemath{\ps}{\,\mathrm{ps}}
\safemath{\meter}{\,\mathrm{m}}
\safemath{\mm}{\,\mathrm{mm}}
\safemath{\cm}{\,\mathrm{cm}}
\safemath{\m}{\,\mathrm{m}}
\safemath{\W}{\,\mathrm{W}}
\safemath{\mW}{\, \mathrm{mW}}
\safemath{\J}{\,\mathrm{J}}
\safemath{\K}{\,\mathrm{K}}
\safemath{\bit}{\,\mathrm{bit}}
\safemath{\nat}{\,\mathrm{nat}}

\safemath{\define}{\triangleq}			% definition

\safemath{\equivalent}{\sim}
\safemath{\distas}{\sim}					% distributed according to
\safemath{\sdiff}{\Delta}				% symmetric set difference

\safemath{\reals}{\mathbb{R}}
\safemath{\positivereals}{\reals_{+}}
\safemath{\integers}{\mathbb{Z}}
\safemath{\posint}{\integers_{+}}
\safemath{\naturals}{\mathbb{N}}
\safemath{\posnaturals}{\naturals_{+}}
\safemath{\complexset}{\mathbb{C}}
\safemath{\rationals}{\mathbb{Q}}

\newcommand*{\fancyrefapplabelprefix}{app}		% Appendix
\newcommand*{\fancyrefthmlabelprefix}{thm}		% Theorem
\newcommand*{\fancyreflemlabelprefix}{lem}		% Lemma
\newcommand*{\fancyrefcorlabelprefix}{cor}		% Corollary
\newcommand*{\fancyrefdeflabelprefix}{def}		% Definition
\newcommand*{\fancyrefproplabelprefix}{prop}		% Proposition
\newcommand*{\fancyrefexmpllabelprefix}{exmpl}
\newcommand*{\fancyrefalglabelprefix}{alg}		% Algorithm
\newcommand*{\fancyreftbllabelprefix}{tbl}		% Algorithm

\frefformat{vario}{\fancyrefseclabelprefix}{Sec.~#1}
\frefformat{vario}{\fancyrefthmlabelprefix}{Thm.~#1}
\frefformat{vario}{\fancyreftbllabelprefix}{Tbl.~#1}
\frefformat{vario}{\fancyreflemlabelprefix}{Lem.~#1}
\frefformat{vario}{\fancyrefcorlabelprefix}{Corr.~#1}
\frefformat{vario}{\fancyrefdeflabelprefix}{Def.~#1}
\frefformat{vario}{\fancyreffiglabelprefix}{Fig.~#1}
\frefformat{vario}{\fancyrefapplabelprefix}{App.~#1}
\frefformat{vario}{\fancyrefeqlabelprefix}{(#1)}
\frefformat{vario}{\fancyrefproplabelprefix}{Prop.~#1}
\frefformat{vario}{\fancyrefexmpllabelprefix}{Ex.~#1}
\frefformat{vario}{\fancyrefalglabelprefix}{Alg.~#1}

\safemath{\dictab}{[\,\dicta\,\,\dictb\,]}

\safemath{\ysig}{\bmy}
\safemath{\ysighat}{\hat{\ysig}}
\safemath{\ysigdim}{M}
\safemath{\xsig}{\bmx}
\safemath{\xsigdim}{N}
\safemath{\nx}{n_x}
\safemath{\zsig}{\bmz}
\safemath{\zsigdim}{\ysigdim}
\safemath{\rsig}{\bmr}
\safemath{\Adict}{\bA}
\safemath{\Adicttilde}{\widetilde{\Adict}}
\safemath{\Adictdim}{\outputdim\times\xsigdim}
\safemath{\avec}{\bma}
\safemath{\avectilde}{\tilde{\avec}}
\safemath{\Bdict}{\bB}
\safemath{\Bdicttilde}{\widetilde{\Bdict}}
\safemath{\Cdict}{\bC}
\safemath{\cvec}{\bmc}
\safemath{\Ddict}{\bD}
\safemath{\Ddictdim}{\ysigdim\times\xsigdim}
\safemath{\dvec}{\bmd}
\safemath{\Ddicttilde}{\widetilde{\bD}}
\safemath{\Bonb}{\bB}
\safemath{\bvec}{\bmb}
\safemath{\Bonbdim}{\ysigdim\times\ysigdim}
\safemath{\noise}{\bmn}
\safemath{\noisedim}{\ysigim}
\safemath{\err}{\bme}
\safemath{\errdim}{\ysigdim}
\safemath{\errset}{\setE}
\safemath{\nerr}{n_e}
\safemath{\delop}{\bP_\errset}
\safemath{\delopc}{\bP_{{\errset}^c}}

\safemath{\cplxi}{\imath}
\safemath{\cplxj}{\jmath}
\safemath{\dict}{\matD}
\safemath{\inputdim}{N}		% number of columns of dictionary D
\safemath{\outputdim}{M}		%number of rows of dictionary D
\safemath{\sparsity}{S}	%sparsity
\safemath{\inputdimA}{{N_a}}	%total number of elements in dictionary A
\safemath{\inputdimB}{{N_b}}	%total number of elements in dictionary B
\safemath{\elemA}{{n_a}}	%number of elements chosen from dictionary A
\safemath{\elemB}{{n_b}}	%number of elements chosen from dictionary B
\safemath{\resA}{\matR_a}	%restriction map to elements of dictionary A
\safemath{\resB}{\matR_b}	%restriction map to elements of dictionary B
\safemath{\subD}{\matS} %subdictionary
\safemath{\subA}{\matS_a} %subdictionary part of A
\safemath{\subB}{\matS_b} %subdictionary part of B
\safemath{\dicta}{\matA} 	% first subdictionary
\safemath{\dictb}{\matB} 	% second subdictionary
\safemath{\hollowS}{H}
\safemath{\hollowA}{H_a}
\safemath{\hollowB}{H_b}
\safemath{\cross}{Z}
\safemath{\coh}{\mu_d}			% coherence dictionary
\safemath{\coha}{\mu_a}			% coherence first subdictionary
\safemath{\cohb}{\mu_b}			% coherence second subdictionary
\safemath{\mubs}{\nu}	%block sub-coherence
\safemath{\cohm}{\mu_m} %mutual coherence
\safemath{\dictset}{\setD}	% set of dictionaries
\safemath{\dictsetp}{\dictset(\coh,\coha,\cohb)}	% set of dictionaries parametrized
\safemath{\dictsetgen}{\dictset_\text{gen}}
\safemath{\dictsetgenp}{\dictsetgen(\coh)}
\safemath{\dictsetonb}{\dictset_\text{onb}}
\safemath{\dictsetonbp}{\dictsetonb(\coh)}

\safemath{\leftside}{U}
\safemath{\rightsideA}{R_a}
\safemath{\rightsideB}{R_b}

\safemath{\indexS}{\setI_S} %set of indices participating in sub-dictionary S

\safemath{\na}{n_a}			% cardinality of set of linearly independent columns of first dictionary
\safemath{\nb}{n_b}			% cardinality of set of linearly independent columns of second dictionary
\safemath{\coeffa}{p_i}	%coefficients for columns of A
\safemath{\coeffb}{q_j}	%coefficients for columns of B
\safemath{\seta}{\setP}		% set of linearly independent columns of A
\safemath{\setb}{\setQ}     % set of linearly independent columns of B
\safemath{\setw}{\setW}	%set of n largest elements of w
\safemath{\setz}{\setZ}	%set of L-n largest elements of z
\safemath{\cola}{\veca}		% generic element of the dictionary A
\safemath{\colb}{\vecb}		% generic element of the dictionary B
\safemath{\cold}{\vecd}		% generic element of the dictionary D
\safemath{\inputvec}{\vecx} 	%coefficient vector (input)
\safemath{\error}{\vece}	%error vector
\safemath{\noiseout}{\vecz} 	%noisy output vector
\safemath{\inputvecel}{x}
\safemath{\inputveca}{\vecx_a}
\safemath{\inputvecb}{\vecx_b}
\safemath{\outputvec}{\vecy}	%output of Dictionary
\safemath{\lambdamin}{\lambda_{\mathrm{min}}}
\safemath{\elltwo}{\ell_2}
\safemath{\ellone}{\ell_1}
\safemath{\ellzero}{\ell_0}
\safemath{\ellinf}{\ell_\infty}
\safemath{\ellinftilde}{\ell_{\widetilde\infty}}
\safemath{\licard}{Z(\coh,\coha,\cohb)}
\safemath{\xsol}{\hat{x}}
\safemath{\xbord}{x_b}		%Solution at the border
\safemath{\xstat}{x_s}		%Solution stationary in l0 prob
\safemath{\xstatLone}{\tilde{x}_s}
\safemath{\order}{\mathcal{O}} %order notation (big O)
\safemath{\scales}{\Theta} %scales as
\safemath{\ones}{\mathbf{1}} %all ones matrix
\safemath{\zeroes}{\mathbf{0}} %all zeroes matrix
\safemath{\thlone}{\kappa(\coh,\cohb)} %treshold l1 problem
\safemath{\constoneA}{\delta} %constant in l1 theorem to save space
\safemath{\constoneB}{\epsilon} %constant in l1 theorem to save space
\safemath{\nlarge}{L}				   %num large elements
\safemath{\sumlarge}{S_\nlarge}
\safemath{\maxlarger}{P_\nlarge}	   % maximum in Gribonval and Nielsen
\safemath{\Pzero}{\textrm{P0}}	
\safemath{\Pone}{\textrm{P1}}
\safemath{\vecfir}{\vecw}			 % \vecv element of the kernel of the dictionary, \vecv=[\vecfir \vecsec]
\safemath{\vecsec}{\vecz}
\safemath{\elvecfir}{w}              % element of vecfir
\safemath{\elvecsec}{z}				 % element of vecsec
\safemath{\nlargefir}{n}
\safemath{\normout}{\gamma}
\safemath{\auxfun}{h}
\safemath{\supp}{\textrm{supp}}%support

\safemath{\indexa}{\ell}
\safemath{\indexb}{r}
\safemath{\indexc}{i}
\safemath{\indexd}{j}

\safemath{\project}{P}%projector

\safemath{\firstslotset}{\setU_1}  % set of UEs for first slot
\safemath{\secondslotset}{\setU_2} % set of UEs for second slot
\safemath{\randomset}{\setS} % generic random scheduling

\safemath{\Tran}{\textnormal{T}}
\safemath{\Herm}{\textnormal{H}}

\newcommand{\orcidlink}[1]{\hspace{1pt}\raisebox{0.5ex}{\href{https://orcid.org/#1}{\includegraphics[height=1.6ex]{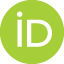}}}}

\newcommand*{\fancyreflstlabelprefix}{lst}
\fancyrefaddcaptions{english}{%
  \providecommand*{\freflstname}{Listing}%
}
\frefformat{vario}{\fancyreflstlabelprefix}{%
  \freflstname\fancyrefdefaultspacing#1#3%
}

\begin{document}

\title{Supervised Device Charting with CSI Measurements\\from Commercial 5G NR User Equipments}
\name{Mischa Vasylyev\orcidlink{0009-0005-4368-6131}, Frederik Zumegen\orcidlink{0009-0006-3046-6321}, Reinhard Wiesmayr\orcidlink{0000-0003-2882-7934}, and Christoph Studer\orcidlink{0000-0001-8950-6267}
    \thanks{We acknowledge NVIDIA for their sponsorship of this research. This work was supported in part by a CHIST-ERA grant for the project CHASER (CHIST-ERA-22-WAI-01) through the SNSF grant 20CH21\_218704.}
    \thanks{Anthropic Sonnet 5 and OpenAI GPT 5.6 Sol were used during preparation of this manuscript. All outputs were independently evaluated and verified by the authors, who take full responsibility for the content of the manuscript.}}
\address{ETH Zurich, Switzerland; e-mail: fzumegen@iis.ee.ethz.ch}

\maketitle

\begin{abstract}
\Gls{RFFI} is a promising approach to distinguish physical wireless devices using hardware-induced signal imperfections. Conventional \gls{RFFI} methods only provide discrete device labels and no human-interpretable representation of the relations among received signals. We propose supervised device charting, which maps location-insensitive \gls{CSI} fingerprints to a low-dimensional chart that visualizes cluster compactness, overlap, and outliers. We evaluate the method with real-world \gls{5GNR} measurements from six commercial smartphones and introduce the \gls{NLER} to quantify class-separation accuracy. Our results demonstrate that two- and three-dimensional device charts provide an interpretable visualization of the learned \gls{RFFI} representation. For three-dimensional device charts, the \gls{NLER} is $0.22\,\%$ for same-day measurements and $7.38\,\%$ for measurements from the next day. The device charts reveal a cross-day distribution shift and map the held-out device close to the known device of the same model. Increasing the device chart dimensions further improves cluster separation at the expense of interpretability.
\end{abstract}

\glsresetall

\section{Introduction}

% DISCLAIMER: all of the authors believe that this method is very likely not practically useful or, at least, needs to be augmented with something that makes it useful

\Gls{RFFI} identifies a wireless transmitter from its radiated RF signal, which is influenced by device-specific hardware characteristics. \gls{RFFI} can thereby complement upper-layer authentication mechanisms~\cite{soltanieh2020review,yan2025radio}. State-of-the-art \gls{RFFI} methods classify a received signal as originating from an enrolled device,\footnote{We say a device is \emph{enrolled} if its fingerprint is known to the \gls{RFFI} system, e.g., when it was included in the training dataset.} reject it as originating from an unknown device, or enroll its previously unseen transmitter \cite{shen2021radio,shen2022towards,xie2021generalizable,mazokha2025mobrffi}. Existing methods, however, output a discrete device label or an unknown-device decision; they provide neither a continuous representation of the relations among samples, such as cluster compactness, overlap, or outliers, nor a human-interpretable visualization. 

\subsection{Contributions}

We propose device charting, where we train a \gls{NN} to map \gls{CSI}-based \gls{RFFI} features from \cite{wiesmayr2025csi} to a continuous human-interpretable representation, i.e., the \emph{device chart}.
Device charting can complement conventional \gls{RFFI} classifiers; a new sample close to one cluster supports the assigned label, whereas a point between clusters or far from all clusters indicates an ambiguous or atypical observation. A network operator can then acquire further \gls{RFFI} samples, request another form of authentication, or recalibrate the \gls{RFFI} system if entire clusters have shifted. In this first work, device-chart distances are not calibrated probabilities and are not used as an automatic acceptance criterion.
Our specific contributions are as follows:
\begin{itemize}
  \item We implement a parametric device charting function with a \gls{NN} trained using a triplet loss and online triplet mining, i.e., we form hard training triplets from each mini-batch on the fly rather than mining them offline beforehand; the trained function maps newly acquired samples without retraining the \gls{NN} or recomputing the existing device chart.
  \item We introduce the \gls{NLER} to measure local class separation accuracy and compare human-interpretable device charts in two or three dimensions against a representation with $D=10$ dimensions.
  \item We evaluate device charting on \gls{5GNR} \gls{PUSCH} measurements published in \cite{wiesmayr2025csi}, which were collected from six \gls{COTS} \glspl{UE}. We demonstrate the efficacy of our approach with \gls{CSI} measured from random locations across two days and with the classification of a held-out \gls{UE} of a known model; we also discuss the relation between the cluster positions and the modem hardware.
\end{itemize}

\subsection{Related Work}

Closed-set \gls{RFFI} methods assign received RF signals to one of the enrolled devices \cite{zhang2021radio,shen2021radio,fu2023radio,wiesmayr2025csi}.
In this work, we extend~\cite{wiesmayr2025csi} to map location-insensitive \gls{RFFI} features inspired by \cite{stephan2025csi} to a continuous device chart instead of assigning a discrete label.

Open-set and metric-learning methods additionally reject unknown devices or support their enrollment and re-identification without retraining the feature extractor \cite{xie2021generalizable,shen2022towards,mazokha2025mobrffi}. Device charting does not implement these decisions, but provides visual context that can supplement them. The open-set method in \cite{ma2025mtpl}, for example, learns a 128-dimensional supervised representation and uses \gls{tSNE}~\cite{van2008visualizing} to visually separate devices. Similar to Sammon's mapping~\cite{sammon1969nonlinear}, \gls{tSNE} embeds a fixed sample set, whereas our \gls{NN} learns a parametric device charting function that maps new \gls{RFFI} samples into the existing chart.

The authors in~\cite{ma2025mtpl} use a publicly available Wi-Fi dataset from~\cite{sankhe2020noradio}, which has fixed transmitter locations, and directly use the I/Q baseband samples as features. In contrast, our \glspl{UE} transmit from randomly varying locations, and our \gls{RFFI} features are designed to suppress location-dependent \gls{CSI} variations.
However, in contrast to our \gls{NN}-based device charting function, classical \gls{tSNE} achieves only low separation accuracy with our \gls{RFFI} features, as shown in \fref{sec:tsne-baseline}.

Separating different physical devices of the same model remains a challenging \gls{RFFI} task \cite{guo2025smorffi}. The closed-set classifier in \cite{wiesmayr2025csi} separates two iPhone 14 Pro units when both are enrolled and classifies both as the same unit if only one unit is enrolled. In this work, we hold out one iPhone 14 Pro unit during training and observe that \gls{RFFI} features from the held-out unit are mapped close to those from the enrolled unit.

Beyond open-set and same-model \gls{RFFI}, current \gls{RFFI} challenges include robustness to temporal variations from the RF channel and the receiver hardware~\cite{soltanieh2020review,yan2025radio}.
To this end, we evaluate the temporal robustness of device charting on measurements from two consecutive days and only use data from the first day for \gls{NN} training.

\section{Supervised Device-Charting Method}

Supervised device charting maps a measured \gls{CSI} sample to a point in a low-dimensional device chart, in which samples from each \gls{UE} form a cluster separated from the clusters of other \glspl{UE}.
We now introduce (i)~location-insensitive \gls{RFFI} feature extraction from~\cite{wiesmayr2025csi}, (ii)~the \gls{NN}-based device charting function, (iii)~its training procedure, and (iv)~the \gls{NLER}.

\subsection{Location-Insensitive CSI Fingerprint Extraction}

Each \gls{CSI} sample in the dataset spans the full bandwidth of 273 \glspl{PRB}, i.e., all $12\cdot273=3276$ active subcarriers, and contains channel estimates from three \gls{DMRS} symbols in one \gls{PUSCH} slot. For the $i$th \gls{DMRS} symbol, $i\in\{1,2,3\}$, we gather \gls{CSI} from four \glspl{ORU} with four receive antennas each in the matrix $\mathbf{H}_i\in\mathbb{C}^{3276\times16}$. Then, we stack the matrices from all three \glspl{DMRS} along the rows.

Inspired by \cite{stephan2025csi}, where the authors \emph{remove} device-related characteristics, we intentionally \emph{extract} such information as our \gls{RFFI} features.
Specifically, we normalize every receive-antenna column to unit norm, compute the compact singular value decomposition, and take its dominant left singular vector. Intuitively, this vector contains, for each subcarrier and \gls{DMRS} symbol, the common part across all distributed receive antennas on the \gls{ORU} side; since the transmitter's single-antenna RF response is present in the channel to every receive antenna, we use this common part as a proxy for the transmitter's RF characteristics.

Finally, we reshape the dominant singular vector to a $3276\times3$ matrix, whose dimensions index the active subcarriers and \gls{DMRS} symbols, respectively, and stack its real and imaginary components along a third dimension. This yields our location-insensitive \gls{RFFI} feature tensor $\mathbf{f}\in\mathbb{R}^{3276\times3\times2}$.

\subsection{Device Charting Function}

We define the parametric device charting function as $g_{\boldsymbol{\theta}}:\mathbb{R}^{M}\rightarrow\mathbb{R}^{D}$, where $M$ is the number of real-valued entries in the \gls{RFFI} tensor and $\boldsymbol{\theta}$ contains the trainable weights of the \gls{NN}. The device charting function maps the vectorized feature $\mathbf{f}$ to one point $\mathbf{z}=g_{\boldsymbol{\theta}}(\mathbf{f})$ in a $D$-dimensional device chart.

We implement $g_{\boldsymbol{\theta}}$ with the two-dimensional convolutional \gls{ResNet} from \cite{shen2022towards,wiesmayr2025csi}. The network implements one input convolution and four residual blocks, each with two convolutional layers. We then apply average pooling, flatten the resulting feature maps, and pass the vector through a fully connected layer with 512 neurons and a dropout probability of $0.2$. Finally, $D$ linear output neurons yield $\mathbf{z}$, i.e., the device-chart point.

\subsection{Supervised Triplet Learning and Online Mining}

We train the device charting function with the supervised triplet loss originally proposed for face recognition in \cite{schroff2015facenet}. To this end, we form each triplet $(\mathbf{f}^{a},\mathbf{f}^{p},\mathbf{f}^{n})$ from the ground-truth \gls{UE} labels: the anchor $\mathbf{f}^{a}$ and positive $\mathbf{f}^{p}$ originate from the same \gls{UE}, whereas the negative $\mathbf{f}^{n}$ originates from another \gls{UE}.

We compute the anchor--positive and anchor--negative distances as $d^{+}=\lVert g_{\boldsymbol{\theta}}(\mathbf{f}^{a})-g_{\boldsymbol{\theta}}(\mathbf{f}^{p})\rVert_2$ and $d^{-}=\lVert g_{\boldsymbol{\theta}}(\mathbf{f}^{a})-g_{\boldsymbol{\theta}}(\mathbf{f}^{n})\rVert_2$, respectively, and use the soft-margin loss $\ln(1+\exp(d^{+}-d^{-}))$ from \cite{hermans2017triplet}.

For every anchor in a mini-batch, we select the farthest same-class sample as the hard positive and the closest different-class sample with $d^{-}>d^{+}$ as the semi-hard negative. If the mini-batch contains no such negative, we select the closest different-class sample as the hard negative.
We train the \gls{NN} with the RMSprop optimizer \cite{tieleman2012rmsprop} using mini-batches of 256 samples. For every mini-batch, the online mining step uses the weights resulting from the previous gradient update.

\subsection{Neighbor Label Error Rate (NLER)}

We now introduce the \gls{NLER} to measure class-separation accuracy.
Let $N$ denote the number of samples in a labeled evaluation set, which comprises the fingerprints and labels $\{(\mathbf{f}_i,y_i)\}_{i=1}^{N}$. To evaluate a trained device charting function, we map each fingerprint to the device chart point $\mathbf{z}_i=g_{\boldsymbol{\theta}}(\mathbf{f}_i)$. For each $\mathbf{z}_i$, we determine $\hat{y}_i$ by majority vote among the labels of its $k$ nearest other device chart points under the Euclidean distance.
We define the \gls{NLER} metric as
\begin{equation}
  \textit{NLER}=\frac{1}{N}\sum_{i=1}^{N}\mathbbm{1}\!\left\{\hat{y}_i\neq y_i\right\}.
  \label{eq:nler}
\end{equation}
Here, $\mathbbm{1}\!\left\{\hat{y}_i\neq y_i\right\}$ equals one if $\hat{y}_i\neq y_i$ and zero otherwise, so $\textit{NLER}\in[0,1]$, i.e., $0\,\%$ to $100\,\%$, with a low \gls{NLER} indicating well-separated \gls{UE} clusters. While the evaluation set typically contains all \gls{UE} classes, we also compute the label-wise \gls{NLER} for each \gls{UE} by restricting both the sum and its normalization to the corresponding $N_y$ evaluation samples.

\section{Results from Real-World Measurements}

We evaluate supervised device charting on real-world \gls{5GNR} measurements via visual inspection and the \gls{NLER}. We now describe the measurement setup and our evaluation protocol. Then, we present two-dimensional device charts, compare them to a \gls{tSNE} baseline, and assess their generalization over time and to a held-out \gls{UE}. Finally, we discuss the trade-off between device chart dimensions and visual interpretability.

\subsection{Measurement Setup}

We use the public CAEZ-5G-DEV-CLASS dataset introduced in~\cite[Sec.~III-C]{wiesmayr2025csi}. The full-stack \gls{5GNR} testbed at ETH Zurich records \gls{CSI} with four \glspl{ORU} placed at the corners of an approximately $4\,\mathrm{m}\times4\,\mathrm{m}$ joint laboratory and office area.

The \gls{5GNR} testbed is compliant with \gls{3GPP} Release 15 and operates at $3.45\,\mathrm{GHz}$ in band N78 with $100\,\mathrm{MHz}$ bandwidth, $30\,\mathrm{kHz}$ subcarrier spacing, and at least one \gls{PUSCH} transmission every $10\,\mathrm{ms}$.

The dataset contains \gls{CSI} from two Apple iPhone 14 Pro units, one in ``gold'' color (\gls{UE} 1a) and one in ``space black'' color (\gls{UE} 1b), an Apple iPhone 16e (\gls{UE} 2), a OnePlus Nord (\gls{UE} 3), a Samsung Galaxy S23 (\gls{UE} 4), and a Google Pixel 7 (\gls{UE} 5). The two iPhone 14 Pro units use a Qualcomm Snapdragon X65 modem, while the remaining models use the Apple C1, Qualcomm Snapdragon 765G, Qualcomm Snapdragon X70, and Samsung Exynos 5300 modems, respectively.

As discussed in \cite{wiesmayr2025csi}, \gls{CSI} from each \gls{UE} was recorded with the same measurement protocol. On the first day, each \gls{UE} was first rotated on a turntable for $30\,\mathrm{s}$, then carried randomly through the measurement area for $60\,\mathrm{s}$, and finally rotated for another $30\,\mathrm{s}$. One day later, each \gls{UE} was only carried randomly through the measurement area for $30\,\mathrm{s}$ after minor changes to the laboratory and the \gls{ORU} setup.
Across all \glspl{UE}, the measurements from the first and second days comprise $83\,619$ and $21\,805$ samples, respectively.

As in \cite{wiesmayr2025csi}, we sort the samples from the first measurement day of every \gls{UE} by timestamp. We use the central $12.5\,\%$, recorded during random motion, as the validation set and the remaining $87.5\,\%$ for training. During training, the validation set controls learning-rate scheduling and early stopping. We use samples from the second measurement day only for testing.

\subsection{Evaluation Protocol}

We train the device charting function on day-one measurements from five \glspl{UE} and hold out \gls{UE} 1b.
We evaluate the trained \gls{NN} on the same-day validation set with the five \glspl{UE}, the next-day test set with the same five \glspl{UE}, and the next-day test set with all six \glspl{UE}.
We show device charts with $D\in\{2,3,10\}$ and evaluate the \gls{NLER} for $k=5$ neighbors.

\subsection{Interpretable Two-Dimensional Device Charts}

\begin{figure*}[t]
  \centering
  \begin{subfigure}[t]{0.32\textwidth}
    \centering
    \includegraphics[width=\linewidth]{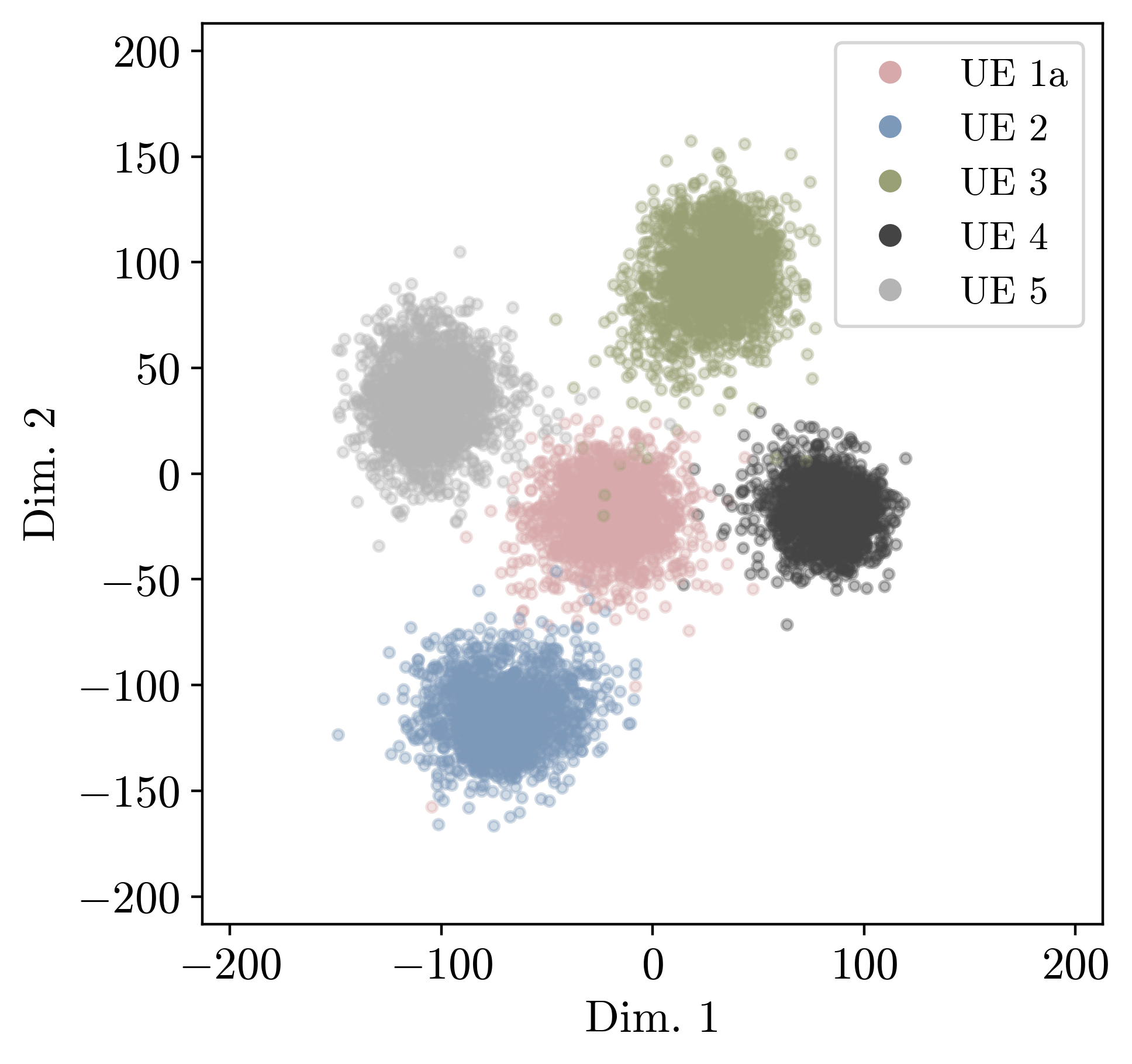}
    \caption{Same day, five enrolled \glspl{UE} (\gls{NLER} $0.55\,\%$).}
  \end{subfigure}
  \hfill
  \begin{subfigure}[t]{0.32\textwidth}
    \centering
    \includegraphics[width=\linewidth]{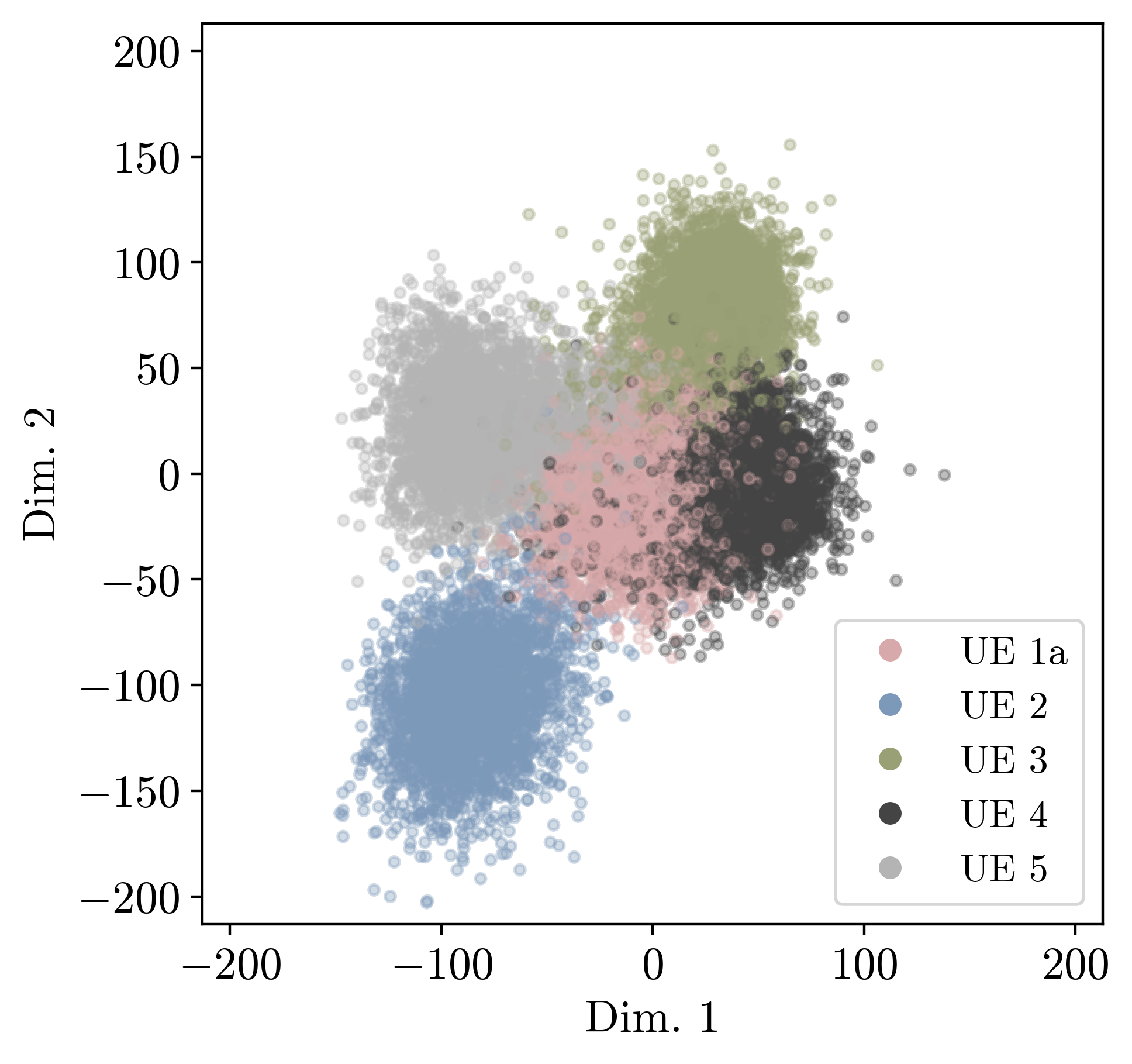}
    \caption{Next day, five enrolled \glspl{UE} (\gls{NLER} $12.94\,\%$).}
  \end{subfigure}
  \hfill
  \begin{subfigure}[t]{0.32\textwidth}
    \centering
    \includegraphics[width=\linewidth]{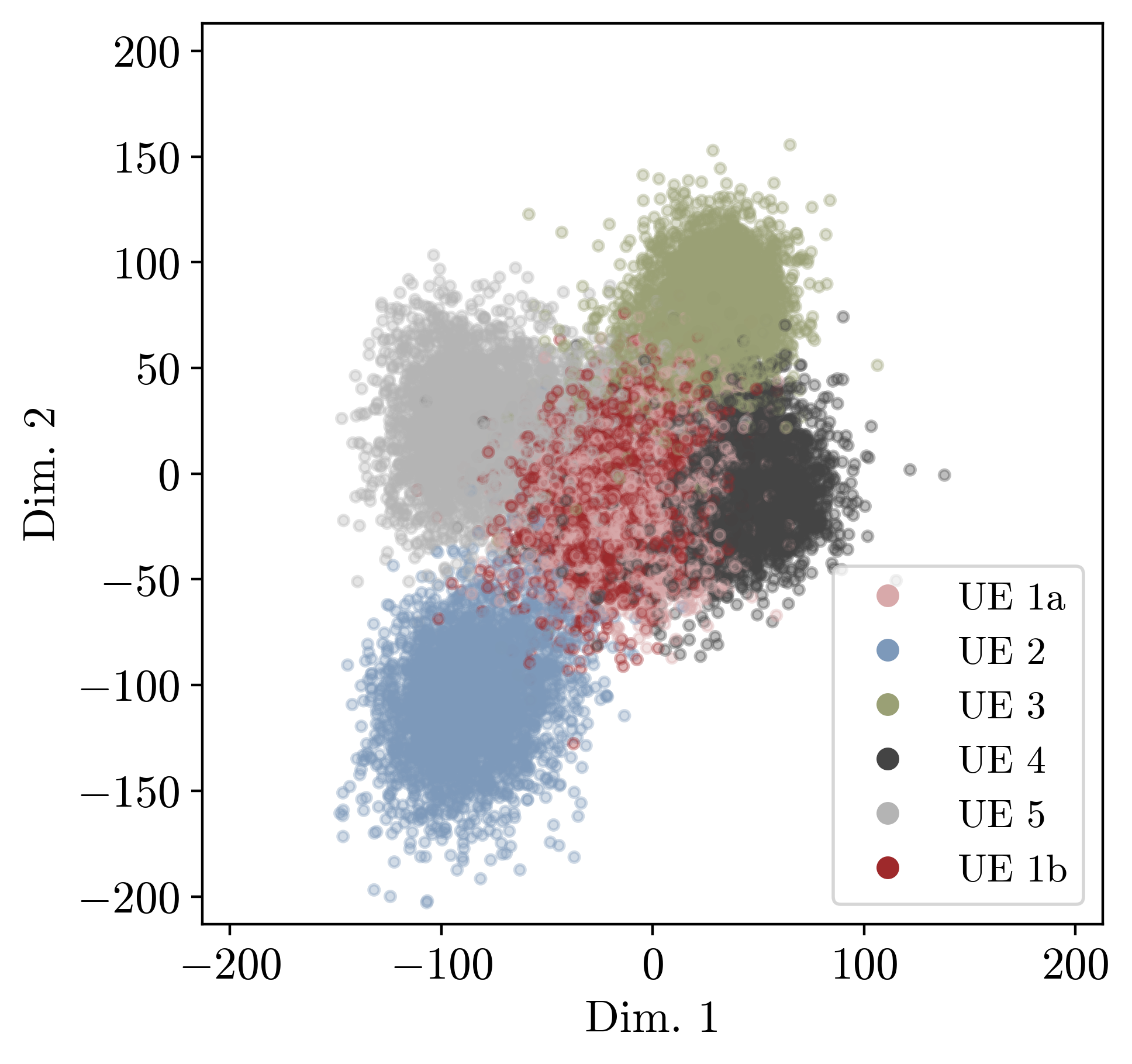}
    \caption{Next day, including the held-out \gls{UE} (\gls{NLER} $26.17\,\%$).}
  \end{subfigure}
  \caption{Two-dimensional device charts obtained with the soft-margin triplet loss. The \gls{NN} is trained on day-one samples from five \glspl{UE}; the ``space black'' iPhone 14 Pro (\gls{UE} 1b) is excluded from training. For the five enrolled \glspl{UE}, the overall \gls{NLER} increases from $0.55\,\%$ on the same day in (a) to $12.94\,\%$ on the next day in (b); (c) shows the placement of the held-out iPhone 14 Pro (\gls{UE} 1b) close to the enrolled unit of the same model (\gls{UE} 1a).}
  \label{fig:charts-2d}
\end{figure*}

\fref{fig:charts-2d}(a) shows the two-dimensional device chart for the same-day validation set with five enrolled \glspl{UE}. The chart forms five visually distinct clusters and achieves an \gls{NLER} of $0.55\,\%$.
We observe that \gls{UE}~1a, \gls{UE}~3, and \gls{UE}~4, which use Qualcomm modems, appear in a common region, whereas the Apple-C1-based \gls{UE} 2 and Exynos-based \gls{UE}~5 form more isolated clusters. Since the triplet loss includes no information about the modems, we report this arrangement as an empirical observation rather than a learned measure of modem similarity.

\subsection{Comparison to a t-SNE Baseline}
\label{sec:tsne-baseline}

As a baseline, we applied standard \gls{tSNE} with $D=2$ and perplexity $50$ to the same-day validation \gls{RFFI} features of the same five enrolled \glspl{UE}.
The \gls{tSNE} baseline reaches an \gls{NLER} of $41.76\,\%$, which by far exceeds the $0.55\,\%$ achieved by our device chart on the same evaluation samples.
We assume that the poor performance of \gls{tSNE} is caused by a residual location-dependence of our \gls{RFFI} features, which the \gls{NN}-based device charting function can learn to suppress.

\subsection{Generalization}

For both generalization experiments, we use the same \gls{NN} trained on day-one measurements from five \glspl{UE}. First, we evaluate generalization over time for the five enrolled \glspl{UE}. Second, we evaluate generalization to the held-out \gls{UE}~1b.

\subsubsection{Generalization Over Time}

\fref{fig:charts-2d}(a) and (b) show two-dimensional device charts with the five enrolled \glspl{UE} from the same-day evaluation dataset and the next-day test dataset, respectively.

We observe displaced and overlapping clusters for the next-day samples in \fref{fig:charts-2d}(b).
\fref{tbl:dimension-nler} also shows that the \gls{NLER} increases from the same to the next day, i.e., from $0.55\,\%$ to $12.94\,\%$ for $D=2$ and from $0.22\,\%$ to $7.38\,\%$ for $D=3$.
The cluster drifts and the higher \gls{NLER} on the next day suggest that the location-insensitive \gls{RFFI} fingerprints still have some residual dependence on the measurement environment and the receiver setup; \cite{wiesmayr2025csi} reports different \gls{UE} positions, minor changes in the laboratory, and different wiring of one \gls{ORU} on the second measurement day.

\subsubsection{Generalization to an Unseen Device}

\fref{fig:charts-2d}(b) and (c) show the two-dimensional device charts without and with the held-out \gls{UE} 1b, respectively. When added to the test set, samples from \gls{UE} 1b are mapped close to the samples from the enrolled \gls{UE} 1a of the same model---exactly as intended.

\subsection{Dimension--Interpretability Trade-Off}

\fref{tbl:dimension-nler} shows the \gls{NLER} for $D\in\{2,3,10\}$. Increasing the dimension from $D=2$ to $D=3$ reduces the \gls{NLER} from $0.55\,\%$ to $0.22\,\%$ for the same-day validation set, from $12.94\,\%$ to $7.38\,\%$ for the next-day test set with five enrolled \glspl{UE}, and from $26.17\,\%$ to $16.67\,\%$ once the held-out iPhone 14 Pro is added to the next-day test set. Increasing the dimension further from $D=3$ to $D=10$ yields diminishing but consistent improvements, reducing the \gls{NLER} to $0.19\,\%$, $7.12\,\%$, and $14.37\,\%$ for the three evaluation sets, respectively.

We can visually inspect $D=2$ and $D=3$, while $D=10$ provides the lowest \gls{NLER} across all evaluation sets; this confirms the trade-off between chart dimension and visual interpretability.

\begin{table}[t]
  \caption{\gls{NLER} for varying device chart dimension $D$ and evaluation datasets.}
  \label{tbl:dimension-nler}
  \centering
  \footnotesize
  \begin{tabular}{@{}lccc@{}}
    \toprule
    Evaluation set & $D=2$ & $D=3$ & $D=10$ \\
    \midrule
    Same day, five \glspl{UE} & $0.55\,\%$ & $0.22\,\%$ & $0.19\,\%$ \\
    Next day, five \glspl{UE} & $12.94\,\%$ & $7.38\,\%$ & $7.12\,\%$ \\
    Next day, six \glspl{UE} & $26.17\,\%$ & $16.67\,\%$ & $14.37\,\%$ \\
    \bottomrule
  \end{tabular}
\end{table}

\section{Conclusions}

We have proposed supervised device charting to map location-insensitive \gls{5GNR} \gls{CSI} fingerprints to a human-interpretable low-dimensional device chart that can be used to complement state-of-the-art \gls{RFFI} methods. To this end, we have adapted the \gls{NN} for discrete device classification in \cite{wiesmayr2025csi} to continuous device charting by replacing its classifier with a linear $D$-dimensional output, and we have trained this \gls{NN} with the supervised triplet loss from face recognition \cite{schroff2015facenet}. The well-separated \gls{UE} clusters show that this combination of \gls{NN} architecture and training loss is well suited for device charting.

Using real-world measurements, we have shown that supervised device charting yields visually separated \gls{UE} clusters and reveals changes between measurement days. The device charting function maps a held-out \gls{UE} close to the enrolled unit of the same model. Increasing the device chart dimensions improves class separation accuracy at the expense of direct visual interpretability. Overall, our results demonstrate the efficacy of the location-insensitive \gls{RFFI} features from \cite{wiesmayr2025csi}. The higher next-day \gls{NLER}, however, suggests that the device charting function remains sensitive to residual effects from changes in the environment and receiver setup.

The triplet labels used in this initial work contain no information about the similarity between different \glspl{UE}. There are several avenues for future work. First, similarity-aware triplet selection could exploit known relations between \gls{UE} models. Second, measurements from more \glspl{UE} and modem families would strengthen our generalization claims. Third, calibrated outlier decisions and integration with open-set enrollment or re-identification are also left for future work.

\balance


\begin{thebibliography}{10}

\bibitem{soltanieh2020review}
N.~Soltanieh, Y.~Norouzi, Y.~Yang, and N.~C. Karmakar,
\newblock ``A review of radio frequency fingerprinting techniques,''
\newblock {\em IEEE Journal of Radio Frequency Identification}, vol. 4, no. 3, pp. 222--233, 2020.

\bibitem{yan2025radio}
G.~Yan, X.~Fu, Y.~Wang, Q.~Zhang, and G.~Gui,
\newblock ``Radio frequency fingerprint identification towards statistical and deep learning features: Review, recent results and future directions,''
\newblock {\em Peer-to-Peer Networking and Applications}, vol. 18, pp. 116, 2025.

\bibitem{shen2021radio}
G.~Shen, J.~Zhang, A.~Marshall, L.~Peng, and X.~Wang,
\newblock ``Radio frequency fingerprint identification for {LoRa} using deep learning,''
\newblock {\em {IEEE} J. Sel. Areas Commun.}, vol. 39, no. 8, pp. 2604--2616, 2021.

\bibitem{shen2022towards}
G.~Shen, J.~Zhang, A.~Marshall, and J.~R. Cavallaro,
\newblock ``Towards scalable and channel-robust radio frequency fingerprint identification for {LoRa},''
\newblock {\em {IEEE} Trans. Inf. Forensics Security}, vol. 17, pp. 774--787, 2022.

\bibitem{xie2021generalizable}
R.~Xie, W.~Xu, Y.~Chen, J.~Yu, A.~Hu, D.~W.~K. Ng, and A.~L. Swindlehurst,
\newblock ``A generalizable model-and-data driven approach for open-set {RFF} authentication,''
\newblock {\em IEEE Trans. Inf. Forensics Security}, vol. 16, pp. 4435--4450, 2021.

\bibitem{mazokha2025mobrffi}
S.~Mazokha, F.~Bao, G.~Sklivanitis, and J.~O. Hallstrom,
\newblock ``{MobRFFI}: Non-cooperative device re-identification for mobility intelligence,''
\newblock {\em arXiv:2503.02156}, Mar. 2025.

\bibitem{wiesmayr2025csi}
R.~Wiesmayr, F.~Zumegen, S.~Taner, C.~Dick, and C.~Studer,
\newblock ``{CSI}-based user positioning, channel charting, and device classification with an {NVIDIA 5G} testbed,''
\newblock in {\em Proc. Asilomar Conf. Signals, Syst., Comput.}, Oct. 2025, pp. 1226--1232.

\bibitem{zhang2021radio}
J.~Zhang, R.~Woods, M.~Sandell, M.~Valkama, A.~Marshall, and J.~Cavallaro,
\newblock ``Radio frequency fingerprint identification for narrowband systems, modelling and classification,''
\newblock {\em {IEEE} Trans. Inf. Forensics Security}, vol. 16, pp. 3974--3987, 2021.

\bibitem{fu2023radio}
H.~Fu, H.~Dong, J.~Yin, and L.~Peng,
\newblock ``Radio frequency fingerprint identification for {5G} mobile devices using {DCTF} and deep learning,''
\newblock {\em Entropy}, vol. 26, no. 1, pp. 38, 2024.

\bibitem{stephan2025csi}
P.~Stephan, F.~Euchner, and S.~ten Brink,
\newblock ``{CSI} obfuscation: Single-antenna transmitters can not hide from adversarial multi-antenna radio localization systems,''
\newblock in {\em Int'l Workshop Smart Antennas}, Sep. 2025.

\bibitem{ma2025mtpl}
Z.~Ma, S.~Fang, and Y.~Fan,
\newblock ``Open-set radio frequency fingerprint identification method based on multi-task prototype learning,''
\newblock {\em Sensors}, vol. 25, no. 17, pp. 5415, 2025.

\bibitem{van2008visualizing}
L.~Van~der Maaten and G.~Hinton,
\newblock ``Visualizing data using {t-SNE},''
\newblock {\em J. Machine Learning Research}, vol. 9, no. 11, 2008.

\bibitem{sammon1969nonlinear}
J.~W. Sammon,
\newblock ``A nonlinear mapping for data structure analysis,''
\newblock {\em IEEE Trans. Computers}, vol. 100, no. 5, pp. 401--409, 1969.

\bibitem{sankhe2020noradio}
K.~Sankhe, M.~Belgiovine, F.~Zhou, L.~Angioloni, F.~Restuccia, S.~D'Oro, T.~Melodia, S.~Ioannidis, and K.~Chowdhury,
\newblock ``No radio left behind: Radio fingerprinting through deep learning of physical-layer hardware impairments,''
\newblock {\em IEEE Trans. Cognitive Communications Networking}, vol. 6, no. 1, pp. 165--178, 2020.

\bibitem{guo2025smorffi}
Z.~Guo, Z.~Jia, J.~Zhu, W.~Huang, and Y.~Chen,
\newblock ``{SMoRFFI}: A large-scale same-model 2.4 {GHz} {Wi-Fi} dataset and reproducible framework for {RF} fingerprinting,''
\newblock {\em Computer Networks}, p. 112309, 2026.

\bibitem{schroff2015facenet}
F.~Schroff, D.~Kalenichenko, and J.~Philbin,
\newblock ``Facenet: A unified embedding for face recognition and clustering,''
\newblock in {\em Proc. IEEE Conf. Computer Vision Pattern Recognition (CVPR)}, Jun. 2015.

\bibitem{hermans2017triplet}
A.~Hermans, L.~Beyer, and B.~Leibe,
\newblock ``In defense of the triplet loss for person re-identification,''
\newblock {\em arXiv:1703.07737}, Nov. 2017.

\bibitem{tieleman2012rmsprop}
T.~Tieleman and G.~Hinton,
\newblock ``Lecture 6.5---{RMSProp}: Divide the gradient by a running average of its recent magnitude,'' Coursera: Neural Networks for Machine Learning, 2012.

\end{thebibliography}
\end{document}